\documentclass[a4paper]{article}

\usepackage{INTERSPEECH2019}
\usepackage{tikz,booktabs,float,color,adjustbox}
\usepackage{multirow}

\title{High Performance Sequence-to-Sequence Model for \\Streaming Speech Recognition}
\name{Thai-Son Nguyen, Ngoc-Quan Pham, Sebastian St\"uker, Alex Waibel}
\address{Institute for Anthropomatics and Robotics, Karlsruhe Institute of Technology}
\email{thai.nguyen@kit.edu}
\begin{document}

\maketitle
\begin{abstract}
Recently sequence-to-sequence models have started to achieve state-of-the-art performance on standard speech recognition tasks when processing audio data in batch mode, i.e., the complete audio data is available when starting processing. However, when it comes to performing run-on recognition on an input stream of audio data while producing recognition results in real-time and with low word-based latency, these models face several challenges. For many techniques, the whole audio sequence to be decoded needs to be available at the start of the processing, e.g., for the attention mechanism or the bidirectional LSTM (BLSTM).
In this paper, we propose several techniques to mitigate these problems. We introduce an additional loss function controlling the uncertainty of the attention mechanism, a modified beam search identifying partial, stable hypotheses, ways of working with BLSTM in the encoder, and the use of chunked BLSTM. Our experiments show that with the right combination of these techniques, it is possible to perform run-on speech recognition with low word-based latency without sacrificing in word error rate performance.
\end{abstract}


\section{Introduction}
Sequence-to-sequence (S2S) attention-based models \cite{chorowski2015attention,chan2016listen} have become increasingly popular for end-to-end speech recognition. Several advances \cite{chiu2018state,zeyer2018improved,weng2018improving,pham2019very} have been proposed to the architecture and the optimization of S2S models to achieve superior recognition performance. In offline scenarios, i.e., batch processing of audio files, the S2S models in \cite{park2019specaugment,nguyen2019improving} have already shown state-of-the-art performance on standard benchmarks. However, methods for employing S2S models in online speech recognition, i.e., run-on recognition with low latency, still needs to be researched, to obtain the desired accuracy and latency.

\cite{jaitly2016online,raffel2017online,chiu2017monotonic} pointed out early that the shortcoming of an attention-based S2S model used in online condition lies in its attention mechanism, which must perform a pass over the entire input sequence for every element of the output sequence. \cite{raffel2017online,chiu2017monotonic} proposed a so-called monotonic attention mechanism that enforces a monotonic alignment between the input and output sequence. Later on, \cite{fan2019online,miao2019online,tsunoo2019online} have addressed the latency issue of bidirectional encoders, which is also an obstacle for online speech recognition. In these studies, unidirectional and chunk-based encoder architectures replace the fully-bidirectional approach to control the latency.

In this work, we analyze the alignment behavior of the attention function of a high-performance S2S model and propose an additional constraint loss to make it capable of streaming inference. By discussing the problems that occurred when adapting a S2S model to be used for a streaming recognizer, we additionally show that the standard beam-search has no guarantee for low-latency inference results, and needs to be modified for providing partial hypotheses. Besides, we argue that the common real-time factor is not a proper choice for measuring the user-perceived latency in online and streaming setup, and propose a novel and suitable technique for the replacement.

In contrast to the earlier works in the literature, our experimental results proved that a bidirectional encoder could be combined with suitable inference methods to produce high accuracy and low latency speech recognition output. With a delay of 1.5 seconds in all output elements, our streaming recognizer can achieve the ideal performance of an offline system with the same configuration. To the best of our knowledge for the first time, a S2S speech recognition model can be used in online conditions without scarifying accuracy.

\vspace{-0.1cm}
\section{Sequence-to-Sequence Model}
\label{sec:model}
In this work, we modify the LSTM-based sequence-to-sequence encoder-decoder model proposed in \cite{nguyen2019improving} to perform high-accuracy online streaming ASR with very low latency. Our model can be decomposed using a set of neural network functions as follows:
\vspace{-0.0cm}
\begin{eqnarray*}
\label{eq:e2easr}
	enc = LSTM(CNN(spectrogram)) \\
	emb = LSTM(Embedding(subwords)) \\
	ctx, attn = SoftAttention(emb, enc, enc) \\
	y = Distribution(ctx + emb)
\end{eqnarray*}

In principle, the functions are designed to map a sequence of acoustic features into a sequence of sub-words and can be grouped into two parts: encoder and decoder. In the encoder, acoustic vectors are down-sampled with two convolutional layers and then fed into several bidirectional LSTM layers to generate the encoder's hidden states $enc$. In the decoder, two unidirectional LSTM layers are used to embed a sub-word unit into a latent representation $emb$. The multi-head \textit{soft-attention} function proposed in \cite{vaswani2017attention} is used to model the relationship between $enc$ and $emb$, which results in a context vector $ctx$. All the functions are jointly trained via the sequence cross-entropy loss by plugging a softmax distribution on top of $ctx$ and $emb$.

As shown in \cite{nguyen2019improving}, this S2S model can achieve highly-competitive offline performance on the Switchboard speech recognition task. However, the model encounters latency issues when being used in online conditions since both, the attention function and the bidirectional encoder network, require the entire input sequence to achieve their optimal performance.


\vspace{-0.2cm}
\section{Streaming S2S ASR}
\label{sec:streaming_s2s}
In this section, we describe our modifications that enable the S2S model to perform online streaming speech recognition with low latency and without loss in performance. The modifications include an additional loss to control the uncertainty of the attention function and search algorithms to infer high-accuracy partial hypothesis.

\vspace{-0.1cm}
\subsection{Discouraging Look-ahead Attention}
\label{ssec:constraint_loss}
\begin{figure}[t]
	\centering
	\includegraphics[width=0.95\linewidth]{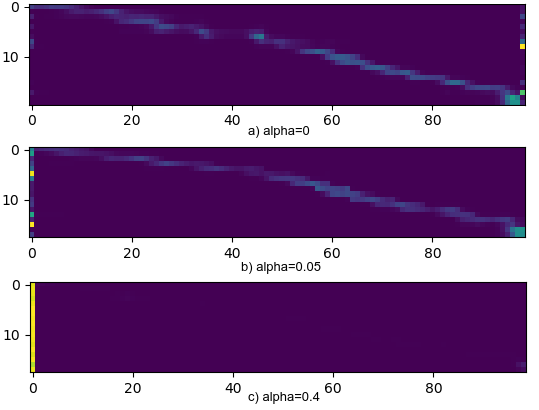}
	\vspace*{-0.3cm}
	\caption{Attention-based alignments provided by a) the regular attention function and b) c) the attention function trained with the constraint loss during the inference of an utterance of 4-seconds length (down-sampling of 4 frames after encoder's layers). The alignments for the tokens 3, 8, 9, 16 are dominated by both start and end frames in a), and dominated by start frames only in b).}
    \vspace*{-0.5cm}
	\label{fig:attn}
\end{figure}

The core of S2S models is the mechanism that autoregessively generates a context vector \textit{ctx} for the prediction of the next token. For the model described in Section~\ref{sec:model}, \textit{ctx} is computed as a sum of all the encoder's hidden states weighted by the \textit{attention scores} which are calculated by the attention function. The attention scores calculated for a specific token typically reveals the positions within the encoder states (or spectral frames) corresponding to the token. So, the attention function can be considered as an \textit{alignment model}. However, this unsupervised alignment does not resemble traditional forced-alignments (or human alignments) in speech recognition. As illustrated in Figure~\ref{fig:attn}a, during a particular inference process, the attention scores produced for many tokens (e.g., \#3, 8, 9, 16) are dominated by the start and end frames, which are not the proper alignments. In this case, the inference still produces the correct transcript, and so the attention function works as it is expected. The mismatch between the attention-based alignment and regular alignment reveals uncertainty that the attention function may have while being optimized with the sequence training likelihood.
Although this uncertainty may not lead to inference errors, the attention function always employs all the encoder's hidden states, which hinders the model from being used in streaming inference. It is preferable for a streaming inference that for the prediction of a token $S$, the attention function only considers past frames until a particular time $S_t$ (the endpoint) and disregards all future frames.

To build such a S2S model for streaming, we investigated the incorporation of an additional loss which discourages the attention function from using future frames during training. Specifically, given token $S$ which belongs to word $W$ in label sequence $L$, we find a region $R_s = (W_t, \infty)$ in which $W_t$ is the end time of $W$ provided by a Viterbi alignment. The attention-based constraint loss is computed as the sum of all attention scores within the region $R_s$ for all $S$ in $L$:
\begin{eqnarray*}
	\mathcal{L}_{attn} = \alpha \sum \limits_{S} \sum_{x}^{|R_s|} Ascore^x_S\\ 
\end{eqnarray*}
The tuneable parameter $\alpha$ adjusts the influence of the constraint loss to the maximum likelihood loss of the label sequence during training. By minimizing both losses simultaneously, we expect that the attention function learns to produce \textit{close-to-zero} scores for the constraint regions for all label tokens while still minimizing the main loss.

\vspace{-0.1cm}
\subsection{Inference for Partial Stable Hypothesis}
\label{ssec:stable_inference}
Beam search is the most efficient approach for the inference of S2S models. Its basic idea is to maintain a search network in which network paths are extended with new nodes with the highest accumulated scores and then pruned away to keep only a set of active paths (or hypotheses).
Typically, the most probable hypothesis for an utterance $X$ is found and guaranteed when the search space constructed from the entire acoustic signals of $X$ is supplied to the search. However, waiting for the complete acoustic signals of $X$ to output inference results is not efficient for a streaming setup. A streaming recognizer must be able to produce partial output while processing partial input. In this section, we describe our search algorithm applied to the proposed S2S model to produce partial output while retaining high accuracy.

Assume that in a streaming setup, at time $t$ we use the proposed S2S model to perform inference for $t$ audio frames. Given a context sequence $C$, the attention function is used to generate $t$ attention scores for the prediction of the next token. We find a time $t_c <= t$ such that the sum of all attention scores from the \textit{covering} window $w = [0, t_c]$ is equal to a constant $\theta = \sum_{x}^{|t_c|} Ascore^x$. When $\theta=0.95$, $w$ covers all dominant attention scores and the context vector generated from $w$ is almost the same as from $[0, t]$. If $t_c$ is observed to be unchanged when $t$ keeps growing, then we consider $t_c$ as the endpoint of $C$. During stream processing, we use a term $\Delta$ to determine if endpoint $t_c$ finally gets fixed as $t_c < t - \Delta$.

We then incorporate the information of endpoints into the beam search to find a partial stable hypothesis. Assume that our beam search can always perform in real-time for $t$ audio frames to produce $N$ considered hypotheses. If all N hypotheses share the same prefix sequence $C$ and the endpoint of $C$ is determined, then we consider $C$ to be an \textit{immortal} part that will not change anymore in the future. When more audio frames are available in the stream, C will be used as the prefix for all search hypotheses, and we repeat this step to find a longer stable hypothesis. Except the condition on endpoints, the idea of finding \textit{immortal prefix} is similar to the partial trace-back \cite{brown1982partial,selfridge2011stability} used in HMM-based speech recognizers.

In addition to the immortal prefix, we also investigated a more straightforward method in which we only consider the \textit{best-ranked} hypothesis and decide on a stable part $C$ based solely on the term $\Delta$. The inspiration comes from the incremental speech recognition approach proposed in \cite{wachsmuth1998integration}.


\vspace{-0.1cm}
\subsection{Bidirectional Encoder}
\label{ssec:bidirect_encoder}
To achieve high performance, bidirectional LSTM have been the optimal choice for the encoder of LSTM-based S2S models. However, due to the backward LSTM, bidirectional LSTM are \textit{not} suited to provide partial and low-latency output as needed for streaming recognizers. The addition of acoustic input will affect all of the encoder's hidden states, which then makes all partial inference results unstable. This effect leads to the fact that stable output can be confidently inferred only when the input is complete. Therefore, earlier works \cite{raffel2017online,he2019streaming,narayanan2019recognizing} switched to unidirectional LSTM in their online models.

In this work, we try to utilize bidirectional LSTM for high-performance speech recognition in a streaming scenario. In the first setting, we investigated the use of the S2S model with a fully bidirectional encoder. First, we train the S2S model to achieve optimal parameters for the offline setup, and with the attention-based constraint loss proposed in Section~\ref{ssec:constraint_loss}. Then, during inference, we update the encoder's hidden states from all available acoustic input before performing the search approaches in Section~\ref{ssec:stable_inference} to find stable hypotheses. As will be shown later, the use of a bidirectional LSTM as this way is possible since the proposed inference methods rely on the determination of endpoints, and the update of encoder's hidden states leads to stabilizing this determination.

In addition to fully bidirectional LSTM, we also experimented with a chunk-based BLSTM approach. During training, we divide input sequences into many non-overlapping blocks of a fixed size of $K$, and then use a BLSTM to compute each block sequentially. To benefit from long-range contextual learning, we initialize the forward LSTM with its last hidden states after processing the previous chunk. The initialization of the backward LSTM can either be a constant or from the previous chunk.
By doing so, the encoder's hidden states can be computed incrementally and efficiently as for unidirectional LSTM. This chunk-based approach is different from \cite{audhkhasi2019forget} and the latency-controlled BLSTM \cite{fan2018online,xue2017improving} that adopt constant initialization of both directions.

\begin{table}[t]
	\caption{WER performance of the S2S model with bidirectional encoder trained with different scales of the constraint loss.}
	\label{tab:constraint_loss}
	\setlength{\tabcolsep}{5pt}
	\centering
	\begin{tabular}{ccccc}
		\toprule
		\textbf{Model} & \textbf{$\alpha$} & \textbf{SWB} & \textbf{CH} & \textbf{Hub5'00} \\
		\midrule
		\multirow{4}{*}{6x1024 BLSTM}
		& & 5.9 & 11.8 & 8.9 \\
		& 0.4 & 6.2 & 12.2 & 9.2 \\
		& 0.2 & 6.1 & 12.1 & 9.1 \\
		& 0.05 & 5.8 & 12.0 & 8.9 \\
		\bottomrule
	\end{tabular}
	\vspace{-0.5cm}
\end{table}

\vspace{-0.2cm}
\section{Experiments}
\label{ssec:setups}
\subsection{Experimental Setup}
Our experiments were conducted on the Fisher+Switchboard corpus consisting of 2,000 hours of telephone conversation speech. The Hub5'00 evaluation data was used as the test set. All the experimental models use the same input features of 40 dimensional log-mel filterbanks to predict 4,000 BPE sub-word units generated with the SentencePiece~\cite{kudo2018sentencepiece} toolkit from all the training transcripts. The models with bidirectional encoder employ six layers of 1024 units while it is 1536 for the unidirectional encoders. We used only 1-head for the attention function in all setups. All models were trained with a dropout of 0.3. We further used the combination of two data augmentation methods \textit{Dynamic Time Stretching} and \textit{SpecAugment} proposed in \cite{nguyen2019improving} to reduce model overfitting. We use Adam \cite{kingma2014adam} with an adaptive learning rate schedule to perform 12,000 updates during training. The model parameters of the 5 best epochs according to the perplexity on the cross-validation set are averaged to produce the final model.

For beam search, we use neither length normalization nor any language model. With a beam size of 8, the experimental models typically achieve their optimal accuracy.

\vspace{-0.2cm}
\subsection{Latency Measure}
\label{ssec:latency_measure}
Neither the commonly used \textit{real-time factor} (RTF) nor commitment latency~\cite{nguyen2020low,li2020towards} are sufficient to measure user-perceived latency for a streaming recognizer. For example, the transcript outputs for an 11-seconds sentence can appear 10 seconds later than a 1-second sentence, but the RTFs measured in two cases can be similar. In this study, we propose to use a different method for measuring streaming latency. Assume that a recognizer processes a sentence S of T seconds in streaming fashion and it outputs N token $s_1$, $s_2$,.. $s_n$ at different timestamps $t_1$, $t_2$,.. $t_n$. And assume the inference time is always a small constant, then timestamp $t_i$ is just when the recognizer is confident of producing $s_i$. The latency of recognizing $S$ with regard to the transcript $s_1$, $s_2$,.. $s_n$ is calculated as the average of all token latencies $t_i \in [1,n]$ normalized by the duration of $S$: $ \sum t_i / (n*T)$. With this measure, the latency of an offline system is always 1 -- as the offline system is only confident for all transcripts until end-of-sentence. In the same way, we simulate the latency of an \textit{instant} recognizer by using a forced-alignment to find $t_i$ for $s_i$.

\label{ssec:latency_blstm}
\begin{table}[t]
	\caption{Latency and accuracy of the S2S model with bidirectional encoder on Hub5'00 test set.}
	\label{tab:latency_blstm}
	\vspace{-0.2cm}
	\setlength{\tabcolsep}{3pt}
	\centering
	\begin{tabular}{ccccc}
		\toprule
		\textbf{Method} & \textbf{Beam Size} & \textbf{$\Delta$} & \textbf{WER} & \textbf{Latency} \\
		\midrule
		\multirow{3}{*}{\textit{Force-Alignment}}
		 & 8 &  & 8.9 & 0.60 \\
		 & 4 &  & 9.1 & 0.60 \\
		 & 2 &  & 9.3 & 0.60 \\
		\midrule
		\multirow{8}{*}{Immortal Prefix}
		& 8 & 20 & \textbf{8.9} & \textbf{0.93} \\
		& 8 & 30 & 8.9 & 0.93 \\
		& 4 & 20 & 9.2 & 0.86 \\
		& 4 & 30 & \textbf{9.1} & \textbf{0.87} \\
		& 2 & 20 & 12.6 & 0.74 \\		
		& 2 & 40 & 10.1 & 0.79 \\
		& 2 & 60 & 9.5 & 0.83 \\
		& 2 & 80 & 9.3 & 0.86 \\
		\midrule
		\multirow{10}{*}{1st-Ranked Prefix}
		& 8 & 30 & 11.2 & 0.75 \\
		& 8 & 50 & 9.6 & 0.80 \\
		& 8 & 70 & 9.3 & 0.84 \\
		& 4 & 30 & 11.3 & 0.75 \\
		& 4 & 50 & 9.6 & 0.80 \\
		& 4 & 70 & 9.3 & 0.84 \\
		& 2 & 10 & 25.8 & 0.62 \\
		& 2 & 30 & 11.4 & 0.75 \\
		& 2 & 50 & 9.7 & 0.80 \\
		& 2 & 70 & 9.3 & 0.84 \\
		\midrule
		\multirow{3}{*}{Combination}
		& 8 & 20-70 & 9.2 & 0.83 \\
		& 4 & 30-70 & \textbf{9.4} & \textbf{0.81} \\
		& 2 & 60-70 & 9.5 & 0.83 \\		
		\bottomrule
	\end{tabular}
	\vspace{-0.5cm}
\end{table}

\vspace{-0.2cm}
\section{Results}
\label{sec:results}
\subsection{Effect of the Constraint Loss}
\label{ssec:baseline_ctc}
In this section, we evaluate the influence of the constraint loss proposed in Section~\ref{ssec:constraint_loss} on the training of the S2S model. We started by using a high value for $\alpha$ and exponentially decreased it to train several systems for comparison. As observed during training, the constraint loss gets small quickly to a stable value depended on $\alpha$. Joint training slows down the convergence of the main loss but does not have a significant impact on the final performance. As shown in Table~\ref{tab:constraint_loss}, WERs are slightly worse with high $\alpha$ and can be similar to the regular training when $\alpha$ is small (e.g., 0.05). Different from that, the constraint loss may largely change the behavior of the attention function. For example, in Figure~\ref{fig:attn}b, the attention function moves the high scores of the mismatched alignment to start frames, instead of end frames as in the regular training. We also found an extreme case when $\alpha=0.4$. The attention-based alignment does not correspond at all to the proper alignment as illustrated Figure~\ref{fig:attn}c. 

Using the model trained with $\alpha=0.05$, we follow the approach in Section~\ref{ssec:stable_inference} to extract the endpoints for all prefixes found during the inference of the evaluation set. We could verify that the extracted endpoints in all sentences match the expectation for streaming inference described in Section~\ref{ssec:constraint_loss}. So we keep this model for further experiments.

\vspace{-0.2cm}
\subsection{Latency on Various Conditions}
Using the S2S model with a bidirectional encoder trained with the constraint loss scale $\alpha=0.05$, we performed several experiments with the inference approaches described in Section~\ref{ssec:stable_inference}. In the experiments, the streaming scenario is simulated by repeatedly feeding an additional audio chunk of 250 ms to the experimental systems for incremental inferences. All the inferences were performed on a single Nvidia Titan RTX GPU, which produced an average RTF of 0.065 with a beam size of 8. The RTF result shows that real-time capacity is not a bottleneck problem in this setup. So we focus on the latency measure proposed in Section~\ref{ssec:latency_measure}.
 
For baselines, we computed the offline WER performance with beam sizes 8, 4, 2, and then used a force-alignment system to produce the \textit{ideal} latency from the offline transcripts. The ideal latency is always 0.6. If we shift the time alignment of the transcripts with 250 ms (i.e., all the outputs have a delay of 250 ms), 500 ms, 1 second, and 1.5 seconds, then we obtained a latency of 0.71, 0.78, 0.86 and 0.91 respectively.

Table~\ref{tab:latency_blstm} presents the accuracy and the latency we achieved when using the \textit{immortal prefix} and \textit{1st-ranked prefix} inference methods with several settings of $\Delta$. Overall, the two methods are consistent with the observations in the HMM-based systems \cite{brown1982partial,selfridge2011stability,wachsmuth1998integration}. Using the \textit{immortal prefix} condition, the final accuracy can be guaranteed as for the offline inference for large beam sizes, e.g., 8 and 4. For a smaller beam size, this condition is not strong enough to deal with unstable partial results -- probably due to the changes of the encoder's hidden states. In the \textit{1st-ranked prefix} approach, increasing  $\Delta$ allows for a flexible trade-off between the accuracy and the latency. The offline accuracy can also be achieved if a very large $\Delta$ is applied.
These results consolidate our findings in two aspects. First, the integration of \textit{$\Delta$} is reliable and crucial for the streaming inferences to work efficiently. And second, the use of the bidirectional LSTM for the encoder is possible and results in high accuracy.

To achieve 8.9\% WER (the offline accuracy), the system needs to delay outputs with an average duration of about 1.5 seconds. To obtain a lower latency of 1 second, the WER increases to 9.2\%, e.g., by using the \textit{immortal prefix} method with $\Delta=20$ and $beam size = 4$. The \textit{combination} of both methods is efficient if we want to reach a latency of 0.81, which is closer to the average delay of 0.5 seconds.

\vspace{-0.1cm}
\subsection{Performance of Different Encoders}
The bidirectional requires additional re-computation of the entire encoder's hidden states for every addition of input signal in the stream. In this section, we investigate two additional network architectures, \textit{unidirectional} LSTM and \textit{chunk-based} BLSTM described in Section~\ref{ssec:bidirect_encoder}, that improve the computational efficiency of the encoder. For chunk-based, we experimented with $K=80$ and $K=200$, as the chunk sizes of 800 ms and 2 seconds. We constantly found that initializing the backward LSTM from the last hidden states of the previous chunk is better than a constant, so we only present the results of this approach. We evaluated two types of encoders in two categories: the best accuracy and the accuracy that the systems retain when maintaining an average delay of 1 second. To do so, we use the same \textit{immortal prefix} inference and experiment with different settings of beam size and $\Delta$.

As shown in Table~\ref{tab:latency_chunkbased}, there is a big gap between the best WER of the unidirectional and bidirectional encoders (12.6\% vs. 8.9\%). The chunk-based encoder, however, is closer to the performance of the bidirectional one when a large chunk size is used. As the encoder's states are fixed early, the inferences are already stable when $\Delta=30$ for all beam sizes. To achieve 1-second delay, all the approaches need to trade-off for an accuracy reduction of 5\% relatively. In term of latency, the chunk-based approach with $K=80$ and $beam size=2$ and $\Delta=30$ is the best setting in this setup.

\begin{table}[t]
	\caption{Latency and accuracy of the S2S models with unidirectional and chunk-based encoders using immortal prefix.}
	\label{tab:latency_chunkbased}
	\vspace{-0.2cm}
	\setlength{\tabcolsep}{3pt}
	\centering
	\begin{tabular}{ccccc}
		\toprule
		\textbf{Encoder} & \textbf{Beam Size} & \textbf{$\Delta$} & \textbf{WER} & \textbf{Latency} \\
		\midrule
		\multirow{5}{*}{Unidirectional}
		& 8 & 30 & 12.7 & 0.94 \\
		& 8 & $\infty$ & 12.6 & 1.00 \\
		& 2 & 20 & 13.6 & 0.82 \\
		& 2 & 30 & 13.2 & 0.85 \\
		& 2 & $\infty$ & 13.1 & 1.00 \\
		\midrule
		\multirow{5}{*}{Chunk-based \textit{K=80}}
		& 8 & 30 & 10.5 & 0.91 \\
		& 8 & $\infty$ & 10.4  & 1.00 \\
		& 2 & 20 & 11.1 & 0.80 \\
		& 2 & 30 & 10.9 & 0.82 \\
		& 2 & $\infty$ & 10.8 & 1.00 \\
		\midrule
		\multirow{5}{*}{Chunk-based \textit{K=200}}
		& 8 & 30 & 10.3 & 0.89 \\
		& 8 & $\infty$ & 10.0  & 1.00 \\
		& 2 & 30 & 11.3 & 0.79 \\
		& 2 & 60 & 10.8 & 0.85 \\
		& 2 & $\infty$ & 10.7 & 1.00 \\
		\bottomrule
	\end{tabular}
	\vspace{-0.5cm}
\end{table}

\vspace{-0.1cm}
\section{Related work}
\label{sec:related_work}
\cite{raffel2017online,chiu2017monotonic} pointed out the problems of the \textit{soft-attention} mechanism on acquiring the entire encoder's states and proposed a trainable \textit{monotonic} attention function to train sequence-to-sequence models for online application. Given a prefix, the monotonic attention function allows finding an encoder position \cite{raffel2017online} or the endpoint of a chunk \cite{chiu2017monotonic} used for prediction of the next token. In our study, we showed that endpoints can also be estimated precisely and efficiently via the regular soft-attention function by controlling its uncertainty.
We further showed that there are more issues to be addressed for high-performance online speech recognition, such as finalizing partial results of the beam search and the use of a bidirectional encoder, and proposed effective methods for addressing theses issues.

\vspace{-0.1cm}
\section{Conclusion}
\label{sec:conclusion}
We have proposed and evaluated several techniques for applying S2S attention-based models to streaming speech recognition. 
Our results show that with these techniques it is possible to produce low latency online recognition results on the Switchboard task without a significant decrease in performance.

\bibliographystyle{IEEEtran}

\bibliography{mybib}

\end{document}